\documentclass[12pt]{article}
  \usepackage{epsfig}
  \usepackage{psfrag}
  \usepackage{a4}
 \newcommand \be {\begin{equation}}                
\newcommand \ee {\end{equation}}
 \newcommand \ba {\begin{eqnarray}}
\newcommand \ea {\end{eqnarray}}
\begin{document}

\title{\bf Can crack front waves explain the roughness of cracks ?}
\author{E. Bouchaud$^1$, J.P. Bouchaud$^2$, D. S. Fisher$^{3,4}$ \\
S. Ramanathan$^5$, J. R. Rice$^{5,6}$}
\maketitle
{\small
\noindent{$^1$ Service de Physique et de Chimie des Interfaces, Centre d'\'Etud
es de Saclay, 
91191 Gif-sur-Yvette Cedex, France \\  
$^2$ Service de Physique de l'\'Etat Condens\'e, Centre d'\'Etudes de Saclay, Orme des Merisiers, 
91191 Gif-sur-Yvette Cedex, France \\
$^3$ Physics Department, Lyman Laboratory, Harvard University, Cambridge, MA 02138, USA \\
$^4$ Division of Engineering and Applied Science, Harvard University,
Cambridge, MA 02138, USA\\
$^5$ Bell Laboratories, Lucent Technologies, Murray Hill, NJ 07974 \\
$^6$ Department of Earth and Planetary Sciences, Harvard University,
Cambridge, MA 02138, USA\\
}}


\begin{abstract}
We review recent theoretical progress on the dynamics of brittle crack fronts and its 
relationship to the roughness of fracture surfaces. We discuss the possibility that the 
intermediate scale roughness of cracks, which is
characterized by a roughness exponent approximately equal to $0.5$, could be caused by the generation, during local instabilities by
depinning, of diffusively broadened corrugation waves, 
which have recently been observed to 
propagate elastically along moving crack fronts. We find that the theory agrees 
plausibly with the orders of magnitude observed. Various consequences and 
limitations, as well as alternative explanations, are discussed. 
We argue that another mechanism, possibly related to damage
cavity coalescence, is needed to account for the observed large scale roughness 
of cracks that is characterized by a roughness exponent approximately equal to $0.8$.
\end{abstract}
\vfill 
\eject

\section{Introduction}

\subsection{Experiments} 

Fracture surfaces are among the best characterized scale invariant objects in
nature \cite{Mandelbrot,EB}: crack profiles have been shown to be self-affine objects,
sometimes over five decades in length scale (from $r=5$ nm to $0.5$ mm). The
roughness exponent $\zeta$ that characterizes 
the typical deviations $\delta h$ of the surface as a function of distance 
along the crack surface $r$ (parallel to the front), as $\delta h \sim r^\zeta$, 
is found to be around 
$\zeta \sim 0.8$.
A typical crack profile is shown in Fig. 1, as an illustration for a rough, self-affine 
object.

Quite surprisingly, the value of $\zeta$ has been 
found to be to a large degree 
{\it universal} \cite{EB1,Roux,EB}, i.e. independent of both the material (glass, metals,
ceramics, etc.) and
of the fracture mode (fatigue, pure tension, stress corrosion, etc.)\footnote{More details on the experimental situations will be provided in section 4} 
More recent experiments, however, have
suggested a more complex scenario, with at least two different 
apparent roughness
exponents \cite{Naveos,Depinning,PRE}: for a given (macroscopic) crack velocity $V_m$, the roughness
exponent for small length scales $r < \xi_c(V_m)$
is found to be around $\zeta=0.5$, whereas for large length scales $r >
\xi_c(V_m)$, the previous value $\zeta=0.8$ is observed. The scale $\xi_c(V_m)$ is a
crossover length which appears to diverge for
$V_m \to 0$ but becomes irresolvably small for the
large $V_m $ that occur in spontaneous dynamic fracture. 
In this interpretation, the value $\zeta \sim 0.5$ 
corresponds to behavior associated with {\it near threshold} crack growth, 
while $\zeta \sim 0.8$ corresponds to `fast' cracks,
for which the effects of the onset are negligible.

As we shall discuss in some detail, it is important to distinguish at least {\it three} 
different roughness exponents \cite{EB}: one describing
the roughness in the direction perpendicular to the crack propagation, a second the roughness in
the direction of the propagation,
and a third one (which we call $\zeta_f$) describing the {\it in-plane} 
roughness of the crack front {\it during its propagation} through the material. 
The exponent characterizing this in-plane roughness, 
which does not directly affect the fracture surface itself, has been
measured by two groups \cite{Sch1,Sch2,EBPD}, on Plexiglas and on metallic alloys respectively. 
These experiments were performed on {\it stationary} crack fronts 
after some growth: a stable crack growth 
geometry was used in \cite{Sch1,Sch2}, and the front was observed {\it in situ} (the sample 
being transparent); more recent experiments
during macroscopically slow growth measured successive crack front
positions in 0.2 s time increments \cite{SchM}. 
In \cite{EBPD}, on the other hand, the fracture was stopped before complete 
failure of the specimen, and the in-plane front morphology was observed after unloading. 
In both sets of experiments, the front roughness index $\zeta_f$ was found to be in the range 
$0.5 - 0.65$, over at least two decades.  

   \begin{figure}
  \psfrag{mean}[ct][ct]{\large $r (nm)$}
  \psfrag{mean-median}[cb][cb]{\large $\delta h (nm)$}
  \centerline{\epsfig{file=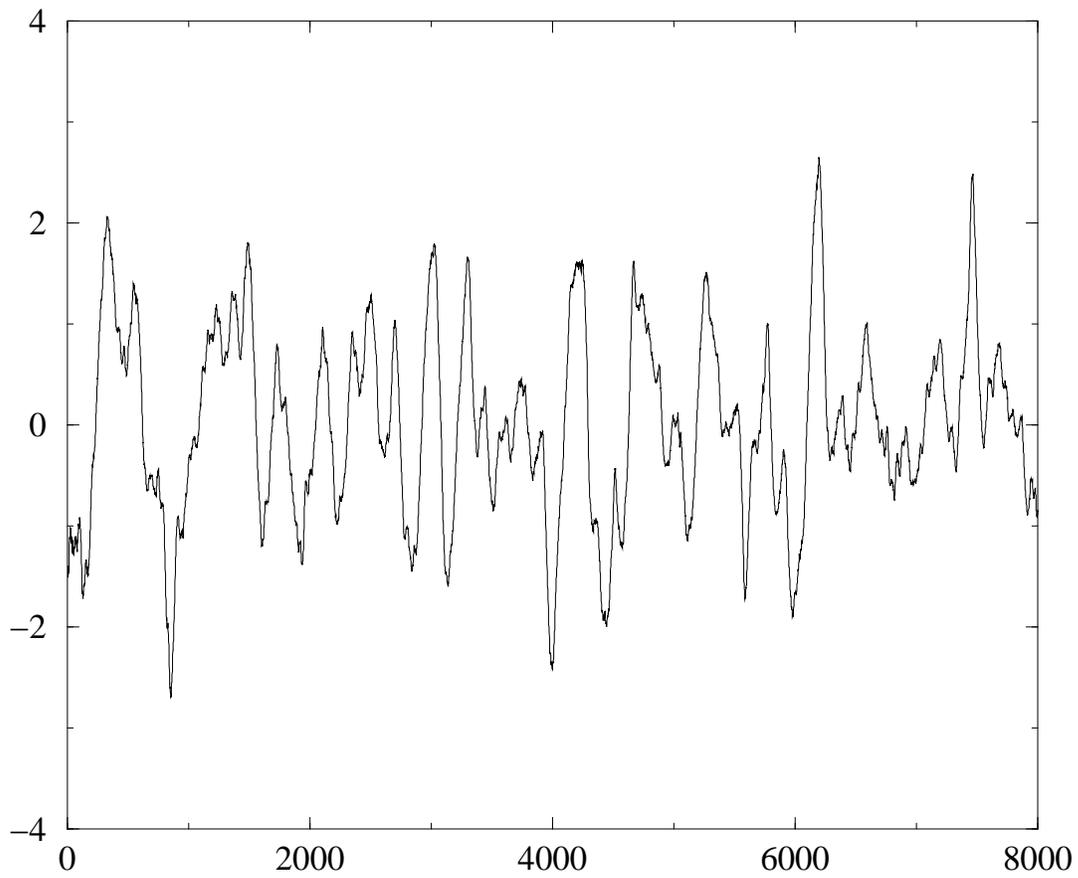,height=0.80\textwidth}}
  \caption{Typical AFM crack profile, measured in a direction perpendicular to the 
crack propagation.
  The material in this case is glass broken under stress corrosion. Horizontal and vertical
  scales are in nanometers. Note that the real slopes are very small.}
  \end{figure}

\subsection{Line models}

Despite numerous recent efforts, there is unfortunately no satisfactory theory that explains the
values of {\it any} of these exponents. 
A {\it qualitatively} useful framework was proposed in \cite{PRL},
in which the crack front
was modeled as an overdamped elastic string moving in a random `pinning' environment
representing the disordered micro-structure of the material. 
This picture provides a natural interpretation for the existence of roughness
exponents, and also
for the appearance of a velocity dependent crossover length separating two
regimes: a ``critical" or ``onset"
regime, appropriate to a crack front just barely able to propagate
through the material, and an ``unpinned" regime, where the front sweeps
through the random pinning at a substantial velocity. 
But this simple elastic-string model is certainly not applicable {\it quantitatively}. 

Refined versions of the crack front model take into account the non-local nature of the
elasticity \cite{JRR1,JRR2,SRVM,Schmitt-Vilotte}.  
Long-ranged elastic effects make the crack front much stiffer  
thereby reducing its roughness 
(and concomitantly that of the fracture surface). 
If elastic waves are ignored, this stiffness, combined with the assumption of only 
short distance correlations in the heterogeneities of the material properties, 
results in a predicted large scale roughness of both the in-plane crack front and 
the fracture surface that grows only {\it logarithmically} with length scale (i.e.
$\zeta=0$) \cite{log-expts,REF}. Although some experiments do indeed observe such weak logarithmic 
roughness \cite{log-expts}, most fracture surfaces appear to be far rougher, at least up to 
a material dependent length scale, beyond which the roughness saturates (or grows much
more slowly \cite{EB}).  

\subsection{Crack front waves} 

But some important physics was left out of these quasi-static calculations: 
the effects of dynamic stress transfer along the crack front caused by elastic waves. 
These were first studied for {\it cracks restricted to a plane}, numerically 
by Morrissey and Rice
\cite{Rice2} and analytically by Ramanathan and Fisher
\cite{RF-planar}. It was found that in an ideal material,  planar distortions 
of the crack front could propagate as waves {\it along} the crack front.  
Such waves would be generated continuously by local variations in the material properties, 
particularly in the critical fracture energy, as the crack advances through a
disordered material. These waves, established in the framework of the full three-dimensional 
vectorial elastodynamics \cite{Mov-Will1} suggested that a 
crack front in an ideal material is even more unstable than had been suggested in the first 3D 
investigations of dynamic cracking through disordered solids by \cite{Rice4} based on a scalar approximation to elastodynamics.  
However, these planar waves only directly affect the in-plane roughness of crack fronts rather 
than the (out of plane) shape of fracture surfaces and, in fact, 
are predicted to be strongly damped whenever the fracture energy is 
substantially velocity dependent \cite{RF-planar}.

Interestingly, another type of crack front wave that {\it can} radically change the roughness 
of fracture surfaces was recently discovered by Ramanathan and Fisher \cite{RF-NP}. Crack fronts 
can indeed also support waves that involve {\it non-planar} deformations of the front, which we
will call {\it corrugation waves} (see also \cite{Mov-Will2} for the perturbative 
elastodynamic solution for a non planar crack).
Although it is not clear at this point whether these waves can propagate forever in an 
ideally elastic material, they can certainly propagate over long enough distances 
to have dramatic effects. Recent observations have indeed shown \cite{fineberg-waves} that 
in glass, perturbations do indeed propagate over long distances. 
These corrugation waves will be reflected in the fracture surfaces. 
Indeed, they are probably the explanation of Wallner lines \cite{Wallner-expts,Hull}, 
the oft-observed 
grooves on the  fracture surfaces of materials that are broken dynamically.

\subsection{Aim of this paper} 

In this paper, we investigate the effects of these waves on the roughness of fracture surfaces, in particular whether 
they might provide a natural
interpretation for the value of the intermediate length-scale roughness exponent of 
$\zeta \sim 0.5$. An analogous effect for the in-plane roughness of the crack front 
was suggested
by Ramanathan and Fisher \cite{REF} but it was not discussed in detail.  
Here we will derive the related result for non-planar crack front deformations, 
and consider both this, and the in-plane case, in a broader context. We will then 
discuss whether 
this scenario is compatible with experimental results and make some predictions about its 
consequences if it indeed is. It is important to stress that the concept of crack waves only
makes sense if the crack is moving sufficiently fast, at least instantaneously. This
might be the case during localized depinning events (see the
recent 
discussion in \cite{SchM}), but is perhaps never justified in the
case of highly ductile, plastic materials where a $\zeta \sim 0.5$ regime is nevertheless 
observed. We discuss in the conclusion alternative models that could explain this value 
of $\zeta$ in the absence of crack front waves.
Finally, we will discuss 
a different mechanism that may be involved in the large length scale, $\zeta \sim 0.8$ regime.

\section{Crack front dynamics}

\subsection{Basic ingredients and notations}

Let us describe the evolution of the shape of the crack front at time $t$ by two functions, $f(x,t)$ 
for its position in the plane of the crack ($f$ is for `front') and $h(x,t)$ for its out of 
plane deformation ($h$ is for `height') with $x$ denoting the coordinate in a direction {\it parallel} 
to the crack front. 
In the following, we choose the $y$ axis 
in the out-of-plane direction and the $z$ axis in the direction of the crack propagation. (Note that these notations differ from those used in, e.g. \cite{Rice2}, where
the coordinate along the crack front is called $z$.)

In an ideal material with no heterogeneities, the front would be straight and 
would propagate in a plane at a uniform velocity $V$ (at least below the 
Yoffe speed) which is a function of the stress intensity factor;
i.e., $f(x,t) \equiv Vt$ and $h(x,t)={\rm constant}$. But in a heterogeneous medium, 
the instantaneous local velocity of the crack front, $V(x,t)$ is constant neither 
in space nor 
in time, and is {\it a priori} very different from the {\it global macroscopic velocity} $V_m$,
because close to the threshold for crack growth, the crack front progresses in a very jerky, 
intermittent manner \cite{RF-onset}. 
 
We will assume a local variation of the material toughness, and therefore of the fracture energy  (the critical energy release rate). 
These heterogeneities affect the dynamics of the crack front in two rather different ways:

\begin{itemize}

\item The variations of the local fracture energy result in a perturbation in the local velocity. 
The resulting change in the shape of the front will modify the stress intensity factor 
\cite{Rice3,Mov-Will1}  
and energy available for fracture at other parts of the crack front thereby affecting the 
crack velocity at these other points.  

\item The {\it direction} of propagation of the crack front can be also 
affected by local heterogeneities. 
Although the details of how this occurs locally will depend on the physics in the process zone 
near the crack front, this should be expected on general grounds: heterogeneities in material 
properties can change the {\it local} loading from being purely tensile (as imposed 
macroscopically) to having a shear component that will tend to make the crack bend in a direction 
that decreases or even cancels 
the Mode II component of the local stress intensity \cite{Cot-Rice,Sethna}. 
On a more microscopic level, the crack may change direction to go around a tougher region that is 
located assymetrically with respect to the local plane of the crack. 
Any assymetric local distortion of the crack front which results from these types of 
heterogeneities will again modify the stress intensity at other points on the crack front, 
in particular by introducing Mode II (and possibly Mode III) components, thereby causing 
non-planar deformations of other parts of the crack front as well.

\end{itemize} 

\subsection{The physical origin of crack waves} 

Because stresses are transferred through the medium and along the crack surface by elastic waves 
-- dilatational, shear and Rayleigh -- the changes in stress intensity factors caused by  
a local disturbance will propagate away from their source at velocities of order the sound speed, 
but the details of this propagation are complicated.  

\subsubsection{In plane waves}

It is instructive to consider what happens to the stress intensity factor along a straight front 
of a planar crack if one small part of it slows down momentarily for some reason, such as an 
encounter with a locally tougher region. The initial changes in the stress at other points 
along the crack front will arrive with the dilatational waves.  Perhaps surprisingly, 
the effects of these will be to {\it increase} the stress intensity factor thereby tending 
to make the other parts of the crack {\it accelerate} rather than decelerate. 
Only after the Rayleigh waves arrive some time later will the stress intensity factor decrease, 
soon becoming less than that before the disturbance arrived and hence tending to slow 
the crack down as one would have expected.  
The crack front waves are a result of the competition between these two effects: a locally tougher 
region will initially cause other parts of the crack to accelerate and then cause them to decelerate. 
In the absence of dissipation, this gives rise to the existence of waves of slowing down 
and speeding up which can propagate in a self-sustaining manner along the crack front at a speed, 
$c_f$ (relative to their source), which is slightly less than the Rayleigh wave speed, $c_R$. 
These carry distortions of the in-plane crack front position $f(x,t)$.

\subsubsection{Corrugation waves}
 
The out-of-plane corrugation waves have a similar origin.  A local distortion of the crack front, 
say in the positive $h$ direction -- `up' --, would be expected to result in 
some Mode II loading at other points on the crack front with a sign which tends to make 
the crack also bend up at these other points thereby keeping the crack front as straight as possible; 
indeed, this is just what the static stress changes due to such a distortion will tend to do.  
But the initial stress changes which arrive with the dilatational waves will have the opposite  effect: 
they carry Mode II stress intensity which tends to make the crack bend {\it down}. 
As was the case for the in-plane velocity changes, this bending effect is negated by the 
Rayleigh waves and at later times the crack will bend in the naively expected `up' direction.  
The competition between these effects of two types of elastic waves, combined with the tendency of 
the crack front to bend so as to cancel the Mode II loading, give rise to propagating waves  
along the crack front and concomitant corrugations in the fracture surface. 
These waves move with a speed,  $c_h$, which is again slightly slower than $c_R$ and 
depends weakly on the overall velocity of the crack front. 

\subsection{An equation of motion for the crack front}

Since we are primarily interested in the fracture surfaces, we will focus on the corrugation waves 
and their effects, returning in section 5.4 to a brief discussion of the in-plane crack front waves. 

A small out-of-plane component of the crack front, $h(x',t')$, will give rise 
to a Mode II stress intensity factor given by \cite{Mov-Will2}:
\be
K_{II}(x,t)=K_I^0 \int_{-\infty}^{\infty} dx' \int_{-\infty}^t dt'\  h(x',t')\ Q(x-x',t-t';V) \label{K-h}
\ee
where the kernel $Q$ is a
 homogeneous function of $x-x'$ and $t-t'$ of degree $-3$ which depends 
on the overall unperturbed velocity of the crack, $V$; and $K_I^0$ is the unperturbed  stress 
intensity factor which we take to be purely Mode I.

We assume that in response to this Mode II local load, 
the crack will tend to bend so as to decrease the Mode II component of the stress 
intensity factor. But, as discussed above, the crack front will also tend to bend in 
response to 
a local assymetry that we parameterize by a random field $\eta({\bf r})$.
A natural assumption with some experimental support is that the crack 
adjusts in such a way that the net Mode II stress intensity factor is zero 
\cite{Cot-Rice,Sethna}:
\be
K_{II}(x,t) - K_I^0 \eta({\bf r}) = 0,\label{eqn-static}
\ee
where $\eta$ is computed at the current position of the 
crack front: ${\bf r}=(x, y=h(x,t), z=f(x,t))$. Therefore, one has, using Eq. (\ref{K-h}):
\be
Q\otimes h = \eta
\ee
everywhere on the crack front and at all times. 
(Here $\otimes$ is the convolution in $x$ and $t$ 
that we wrote explicitly in Eq.(\ref{K-h})). 

We can invert this to find the response of the crack to a local bending heterogeneity:
\be
h = P \otimes \eta \qquad P\equiv Q^{-1} \label{def-P}
\ee
The `propagator' $P(x-x',t-t')$ is a complicated homogeneous function of degree $-1$ 
which is only known 
explicitly as an integral expression; it includes the effects of all of the three types 
of elastic waves.  

The above equation (\ref{eqn-static}) assumes that the crack front `follows' perfectly
the local randomness. It might be interesting to generalize this equation to describe
the fact that the crack will `react' to a change of stress intensity factor with a 
certain lag. Thus we propose an effective equation of motion for the direction of 
the crack front of the following form \cite{REF,RF-NP} :
\be
\frac{\partial^2 h(x,t)}{\partial t^2}=
\frac{V^2}{\ell_r}\biggl(-\frac{K_{II}(x,t)}{K_I^0} + 
\eta({\bf r}) \biggr)    \label{eqn-motion}
\ee
where $\ell_r$ is a microscopic `adaptation' length.  
Intuitively, Eq. (\ref{eqn-motion}) means that the local orientation angle of the surface, 
equal to $\partial h/\partial z$ changes at a rate proportional to $K_{II}/\tau_r$, with
$\tau_r=\ell_r/V$. Since the 
crack is moving at a velocity $V$, derivatives with respect to $z$ are, for a weakly perturbed
crack front, simply related to time derivatives through $\partial z=V \partial t$.
On length scales large compared to $\ell_r$, one can set the left hand side of equation 
(\ref{eqn-motion}) to zero, and recover (\ref{eqn-static}).  In the following, 
we will assume that the length $\ell_r$ is microscopic.

\subsection{The corrugation waves: diffusion and dispersion}

The corrugation waves of the crack front arise from a zero in the Fourier transform of $Q$ at a real 
(or almost real, see below) value of $\omega/|q|=s_h$ which gives rise to a divergence of $P$ for $(x-x')=\pm s_h(t-t')$ 
where
\be
s_h=\sqrt{c_h^2-V^2}
\ee
is the speed of the corrugation waves in the direction parallel to the moving crack front (Note that the 
total speed $c_h$ relative to the source is indeed given by $c_h^2 = s_h^2 + V^2$).

The wave speed $c_h$ is found to depend weakly on the crack front velocity $V$:
it varies from $0.96\, c_R$ when $V \ll c_R$ to $c_R$ as $V \to c_R$. 
The numerical
solution suggests that $c_h$ has a very small but non zero imaginary part
$\epsilon(V) c_R$, with $\epsilon$ of
order $10^{-4}$ for small $V$ to $2 \, 10^{-3}$ for $V = 0.6 c_R$
\cite{RF-NP}. This means that, strictly speaking, corrugation waves 
will not propagate indefinitely. Since $\epsilon$ is so small, it is
a reasonable approximation to ignore, at least for now, its effects.

The behavior near to this singularity of $P$, dominates the long time behavior of the crack front.  
The important parts have the form:
\be
P(x,t) \propto \bigl(\frac{C_b c_s}{x-s_ht} -\frac{ C_b c_s}{x+s_ht} \bigr) \label{prop}
\ee
(in contrast to a sum of delta-functions at $x\pm s_h t$ for 
conventional waves). 
Here $C_b$ is a velocity dependent dimensionless numerical coefficient, and $c_s$ 
is the shear wave speed.
[Note that the dimension of $P$ is $[T]^{-1}$, as it should be 
since $\eta$ is dimensionless in 
Eq. (\ref{def-P}), and convolution brings an extra $[L][T]$ factor.]
The primary effects of a perturbation propagate away from its source in 
two directions that are at an angle 
$\arccos(V/c_h)$ from the direction of propagation of the crack -- i.e., almost 
{\it parallel} 
to the crack front for a slowly advancing crack. 

From Eq.(\ref{prop}), the shape of any disturbance would
persist without broadening for arbitrarily long times (for $\epsilon =0$). 
But this result holds only for a perfect elastic medium with an infinitely sharp crack and no lag in response to 
bending forces (i.e., $\lambda_r =0$).  In reality, the two sharp peaks of Eq. (\ref{prop})
will be broadened by various mechanisms. The first one, discussed by Ramanathan and Fisher
\cite{RF-planar}, is the existence of Kelvin-like viscoelastic effects, i.e. a delay time 
$\tau_d$ between stresses and strains.\footnote{This effect was not treated correctly in 
reference \cite{RF-planar} due to the nature of the boundary conditions 
in the moving frame of the crack front.  For the case of in-plane cracks discussed there, 
a detailed calculation has been carried out \cite{DSF-unpub} and leads to qualitatively similar results.  
In principle, this could also be done for the out-of-plane dynamics although the technical details are 
likely to be exceedingly cumbersome.}
This will lead to a diffusive-like spreading of the peaks, with a diffusion constant of the order
of $D_d \sim c^2 \tau_d$, whose actual value will 
involve many details of the relaxation processes as embodied in the frequency dependence of the elastic moduli 
\cite{DSF-unpub}. (In the above formula and in the rest of this paper, $c$ denotes a typical wave propagation speed 
-- as far as orders of magnitude are concerned, we do not need to distinguish between the different wave speeds.)  

Two other effects are caused by the heterogeneities in the medium in
which the elastic waves and the crack front propagate. These inhomogeneities can scatter the (bulk or 
Rayleigh) elastic waves which mediate the crack front dynamics and thereby give rise to broadening of 
the ideal crack front waves. But in addition, the small scale corrugations in the fracture surface, 
created by the heterogeneities and propagated by the crack front waves themselves can act as disorder to 
scatter the longer wavelength crack front waves. (This may lead to interesting non-linear 
feedback effects, see \cite{fineberg-waves}).  The simplest expectation is that both of these will induce diffusive like 
spreading of the peaks of Eq. (\ref{prop}) (see e.g. \cite{Claudin}). If the scattering were strong, one would
expect the corresponding diffusion constant to be of order $D_s=c \ell_s$,
where $\ell_s$ is the correlation length of the relevant inhomogeneities. 

These effects together 
give rise to a change of the important singular parts of the Fourier transform of the propagator from the ideal case
corresponding to  Eq.(\ref{prop}) which in Fourier space is:
\be
\hat{P}(q,\omega) \sim \frac{C_b c_s}{\omega -s_h|q|}+ \frac{C_b c_s}{-\omega-s_h|q|}
\ee
to
\be
\hat{P}(q,\omega) \sim \frac{C_b c_s}{\omega -s_h|q|+iDq^2}+ \frac{C_b c_s}{-\omega-s_h|q|-iDq^2}  . \label{prop-D}
\ee
where $D=D_s+D_d$. The shift of the pole by $i Dq^2$ indeed corresponds to a diffusive damping term of the 
form $\exp[-Dq^2 (t-t')]$ in the time domain.

Physics within the process zone will also affect the propagation of disturbances along the crack front.   
One might expect that the non-instantaneous response (i.e. $\ell_r \neq 0$) 
of the crack growth direction embodied in the equation of motion, 
Eq. (\ref{eqn-motion}), would give rise to similar diffusive-like broadening.  But in fact, 
this primarily gives rise to {\it dispersion} of the waves; this term
in Eq. (\ref{eqn-motion}) yields, in Fourier space, 
$-\ell_r/V^2 \omega^2 \hat h(q,\omega)$. 
Since this contribution must be added to $Q$, one finds that near the poles (i.e. when $Q$ is small), 
the propagator $\hat P$ can be written as:
\be
\hat{P}(q,\omega) \sim \frac{C_b c_s}{\omega -s_h|q|-Jq^2+iDq^2}+ \frac{C_b c_s}{-\omega-s_h|q|-Jq^2-iDq^2}, \label{prop-F}
\ee
which is valid when both $q$ and one of $\omega \pm s_h|q|$ are small. Here, the coefficient $J$ that describes
dispersion effects is given by $J = C_b c_s s_h^2 \ell_r/V^2$. As will be clear below, however,
dispersion effects will not drastically affect the roughness statistics.  

Fourier transforming Eq. (\ref{prop-F}) gives rise to the final form of the propagator, which is a sum of 
a right moving contribution $P^R(x,t)$ that depends on $\chi_R = x - s_h t$ and a 
similar left moving contribution $P^L(x,t)$
that depends on $\chi_L = x + s_h t$. We find:
\ba
P^R(\chi_R) & \approx & C_b c_s \int_0^\infty \frac{dq}{\pi} \sin(q\chi_R +Jq^2t)e^{-Dq^2t}  \\
 & \approx & \frac{C_b c_s}{2 \sqrt{\pi Dt}} 
\Im \biggl[\exp(- \tilde \chi_R^2)\ {\rm erfc}(-i \tilde \chi_R)\biggr]
, \quad{\mbox {with}}\quad
 \tilde \chi_R \equiv \frac{\chi_R}{2\sqrt{(D+iJ)t}}.\nonumber\label{prop-right}
\ea
In the above equation, ${\rm erfc}$ is the complementary error function, and $\Im$ denotes the imaginary part.
The above result holds for $|\chi_R| \ll s_h t$ since we have used the expression of the propagator 
Eq. (\ref{prop-F}) which is only valid close to the pole.  

For $J\ll D$, (\ref{prop-right}) is an antisymmetric function of $\chi_R$ which decays as $1/\chi_R$ for $\chi_R \gg \sqrt{Dt}$ (in agreement with 
Eq. (\ref{prop}) above), vanishes linearly for 
$\chi_R\ll \sqrt{Dt}$, and has a peak (trough) for 
$\chi \sim \pm \sqrt{Dt}$. This function is plotted in Fig. 2. Interestingly, this shape 
is similar to what has been observed in \cite{fineberg-waves} .

  \begin{figure}
  \psfrag{mean}[ct][ct]{\large $\tilde \chi$}
  \psfrag{mean-median}[cb][cb]{\large $\sqrt{Dt} P(\tilde \chi)$}
  \centerline{\epsfig{file=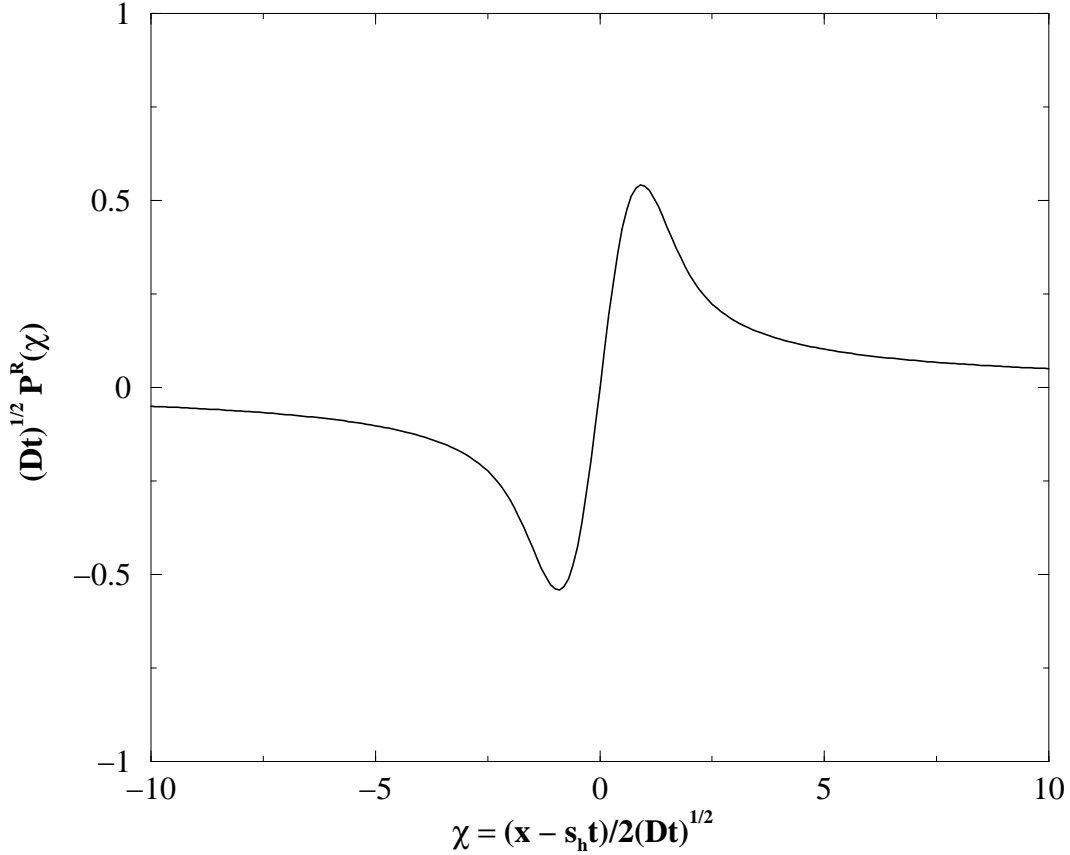,height=0.80\textwidth}}
  \caption{Shape of the rescaled propagator $P(\tilde \chi)$ 
  as a function of the rescaled variable 
  $\hat \chi= (x \pm s_h t)/2\sqrt{Dt}$. Here we have set $C_b c_s/
  2\sqrt{\pi}=1$.}
  \end{figure}

\section{Corrugation wave mediated roughening}  

\subsection{Quantities of interest} 

The dependence of the random local bending tendency, $\eta$ (see Eq. (\ref{eqn-motion})), 
on $h$ does not play 
an important role, and we assume a simple form for its random dependence on $x$ (the direction along the crack front)
and $z$ (the direction parallel to crack propagation): Gaussian with mean zero and covariance given by
\be
\langle \eta(x,z)\eta(x',z') \rangle = \sigma^2 {\cal G}\left(\frac{(x-x')^2+(z-z')^2}
{\xi_0^2 }\right)
\ee
with $\sigma$, the dimensionless root mean square amplitude of the random bending, 
$\cal G$ a certain short range function and $\xi_0$ the correlation length of this randomness.
In a disordered material, there are a priori many different length scales associated to
different types of heterogeneities: size of precipitates, microcavities, metallurgical
grains, quenched in stresses, etc. The relevant heterogeneities will actually depend 
on the observation scale. In the following, for simplicity, we will assume that we are
only interested in length scales large compared to $\xi_0$, and replace the function 
$\cal G$ by a (two dimensional) $\delta$-function. This might however not always
be justified (see section 5.4 and \cite{Schmitt-Vilotte}).

The existence of crack front waves means that a local
variation of the material properties that is anisotropic or located just off the plane of the crack, 
say near $(x_0,z_0)=(x_0,Vt_0)$, will result in a perturbation $h(x,t)$ 
of the deviation of the crack from planar which propagates, relative to 
$(x_0,z_0)$, at a velocity $c_h$. Using the results of the previous section, the 
perturbation induced by a variation $\eta(x_0,z_0)$ can be written as \cite{Rice2,Rice3}:
\be
h(x,t) =   \int_{-\infty}^{+\infty} dx_0 \int_{0}^{t} dt_0
\left[P^R+P^L\right](x-x_0,t-t_0) \  \eta(x_0,z_0=Vt_0) \label{0},
\ee
where $t=0$ is the time at which the front penetrates into the disordered
region. Again, we have assumed that the perturbation from a straight front is small in order to
replace $z_0$ by $Vt_0$. 

From this expression, one can compute the correlation function of the fracture surface heights
from the function:
\be
B(r_x,r_z)= \left \langle [h(x+r_x,z+r_z)-h(x,z)]^2 \right \rangle,
\ee
that is often measured measured experimentally. The brackets refer to an average 
over the point $(x,z)$, which 
--- provided the measurements are taken in a region sufficiently far from where the crack 
front enters the random heterogeneities and starts to roughen, and sufficiently small
so that the crack does not accelerate --- can be
replaced by an average over the randomness.  The roughness exponent $\zeta$, is defined by  
$[B({\bf{r}})]^{1/2} \sim |{\bf{r}}|^{\zeta}$, and may in general depend on the direction of 
${\bf{r}}$.  
The roughness of the fracture surface reflects the temporal history of the non-planar 
deformations of the crack front.  

\subsection{Roughness correlation function}

It is instructive to consider how a given `asperity' (i.e. a given local variation of the material
properties) --- located at $(x_0,z_0)$ --- 
contributes to the roughness correlations.   The dominant effects of this asperity will be carried by the 
left and right moving corrugation waves, with a diffusive-like spreading of these waves in time. 
Points outside of these spreading waves will not be affected appreciably.  
If the two points of interest ${\bf {r}},{\bf {r'}}$ are both affected by the waves, 
the resulting deformations 
of the  crack front at the two points are highly 
correlated and do not contribute much to $B({\bf r)}$ unless:
\begin{itemize}
\item either their separation in time, $|r_z/V|$, is comparable or greater than the time, 
$(z-z_0)/V$, 
since the waves left the asperity; 
\item or their separation perpendicular to the direction of propagation,  $|r_x-s_hr_z/V|$ 
for the right-moving waves, is comparable to or greater than the diffusive spreading, $\sqrt{Dr_z/V}$. 
\end{itemize}
Thus the dominant contributions to the mean-square height differences will be from asperities which are 
{\it within} a parabola opening backward from one of the points with its axis along the wave direction, 
but {\it not within} the similar parabola opening backward along 
the same wave 
direction from the other point (see Figure 3).  

   \begin{figure}
  \centerline{\epsfig{file=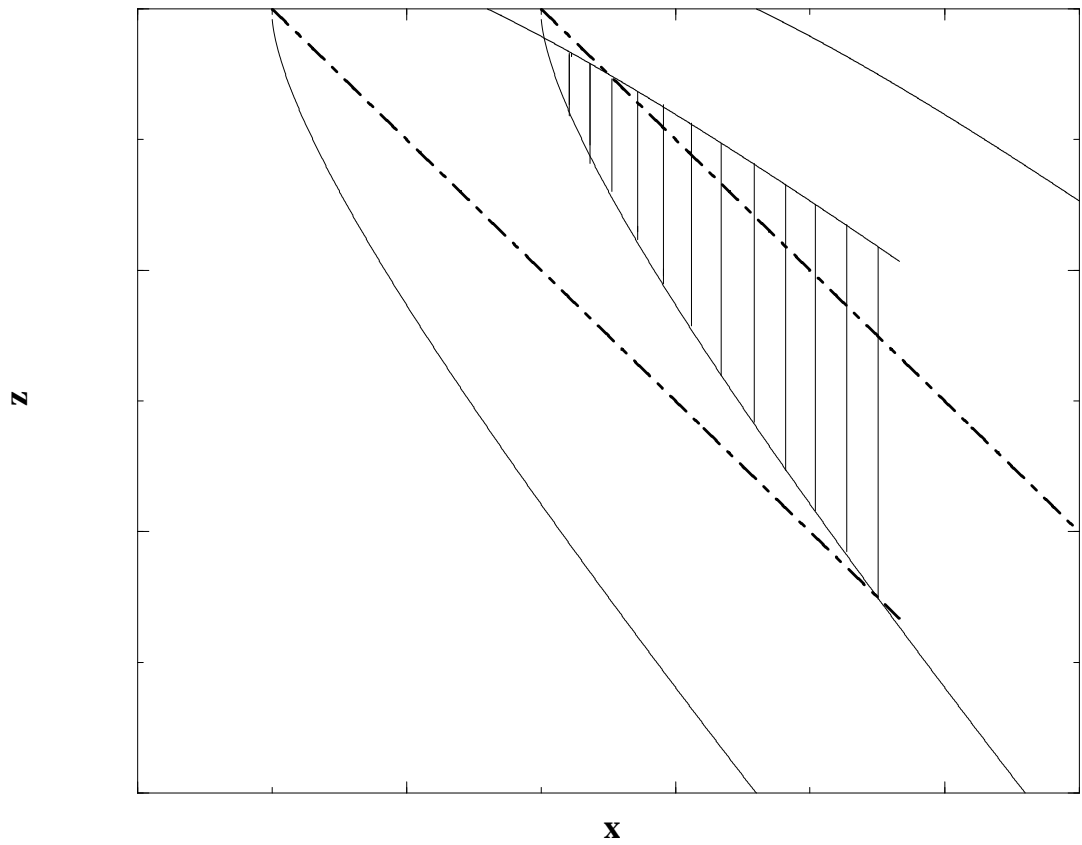,height=0.80\textwidth}}
  \caption{The contribution to the difference of roughness between two
given points comes from asperities within parabolas opening backwards 
along the wave propagation direction, but not from their common 
intersection (hatched region).}
  \end{figure}

The mean-square roughness is approximately given by a sum of two terms, 
one from the right-moving and the other from the left-moving waves. 
The effects of the cross-terms between the right and left moving waves are small, because they carry
signals that come from uncorrelated asperities. We find that: 
\be
B({\bf r}) \approx \frac{4C_b ^2 \sigma^2 \xi_0^2 c_s^2}
{V}\left[{\cal F}(|r_x-s_hr_z/V|,r_z) + 
{\cal F}(|r_x+s_hr_z/V|,r_z)\right] 
\ee
where the function ${\cal F}$ has the following asymptotic behaviour:
\be 
{\cal F}(|\chi|,r_z) \approx \frac{|\chi|}{2D}
\ee
for $|\chi| \gg \sqrt{Dr_z/V}$ and
\be
{\cal F}(|\chi|,r_z) \approx \sqrt{\frac{r_z}{\pi VD}}\  \frac{\cos \theta/2}{\cos^2 \theta},
\ee
where $\tan \theta=J/D$, for $|\chi| \ll \sqrt{Dr_z/V}$.

Let us comment these results. First consider the roughness measured {\it perpendicularly} to the direction of
crack propagation, i.e. along the $x$ direction, which corresponds to $r_z =0$. We find that the roughness is given by:
\be 
\delta h(r_x) \sim \sigma \xi_0 c_s \sqrt{\frac{r_x}{DV}}.\label{result-x}
\ee
Note that if the distance traveled by the crack since it entered the disordered region is finite and equal to
$L (=Vt)$, the above result is only valid if $r_x \ll \xi_L$, with $\xi_L=\sqrt{DL/V}$. For longer distances,
the roughness saturates in this regime.

In the direction {\it parallel} to the crack propagation --- i.e. perpendicular to the crack front, $r_x=0$  --- the roughness is given by
\be
\delta h(r_z) \sim \sqrt{\frac{s_h}{V}}\ \  \sigma \xi_0 c_s \sqrt{\frac{r_z}{DV}},\label{result-z}
\ee
i.e the roughness is reduced by a factor of order $\sim \sqrt{c/V}$ compared to that in the $x$ 
direction. This difference in the amplitudes in the 
two directions should be very pronounced at low crack growth velocities.

Another notable effect of roughness due to 
corrugation waves is that along the direction of 
propagation of the crack front waves, i.e. for $r_x = 
s_h r_z/V$, one expects to observe ridges and grooves oriented in these directions.  
The long-distance roughness is somewhat suppressed in the wave 
directions where one finds:
\be
\delta h \sim \sigma \xi_0 c_s \left(\frac{r_z}{V^3 D}\right)^{1/4},
\ee
which is a factor $(VD/c^2r_z)^{1/4}$ smaller than in the direction parallel to the crack propagation, where
it is already reduced compared to that in the $x$ direction. Interestingly, 
if such a reduced roughness direction is observed it could be used to 
{\it determine} the 
local velocity $V$ at which the crack was propagating \cite{fineberg-waves}.   

Note that the above results for the roughness exponents obtained here ($\zeta=1/2$ or $1/4$ 
in a particular direction) are, largely coincidentally, 
identical to that for an over-damped elastic line driven through random impurities (the so called
Edwards-Wilkinson model \cite{BS}). This is in spite of the fact that the the elasticity of a crack front is non
local (i.e., it is described by a $|q|$ wave-vector dependence of the energy, rather than the
$q^2$ dependence of the oversimplified string model). This non local elasticity 
is in a sense responsible for the
existence of crack waves; however, the diffusive spreading of these waves is
eventually the dominant factor that determines
the statistics of the crack roughness; this is the cause for the equivalence of the 
roughness exponents to that of the elastic string with diffusive dynamics.

So far, we have ignored the effects of the small imaginary part $\epsilon(V)
c_R$ of the corrugation wave velocity. At sufficiently long times, this will
damp the waves more rapidly than diffusively. The decay of the effects 
of a localized perturbation will eventually change from the $1/\sqrt{Dt}$
decay of the peaks in $h$ (see Eq. (10)) to $1/\epsilon(V)
c t$ at longer times. But the crossover time will be of order 
$t_\epsilon \sim D/\epsilon^2 c^2$. In directions parallel to the crack front
the roughness will thus only be reduced on length scales larger than
$\xi_\epsilon \sim D/c \epsilon^2$ which is macroscopic even if $D/c$ is 
nanometric. Perpendicular to the crack front, the reduction of the roughness
will occur on scales smaller by a factor $V/c$. On asymptotically long
length scales, the predicted roughness would be reduced to its logarithmic 
form found in the absence of crack front waves but with an amplitude 
increased by a factor $1/\epsilon(V)$.
 
\section{Comparison with experiments}

In this section, we will compare the above predictions with several experimental
observations on materials as different as glass, intermetallic-based or metallic-based
alloys, fractured in stress corrosion, fatigue or pure tension. Our main prediction 
(Eq. (\ref{result-x})) concerns the amplitude 
and form of the roughness correlations in the regime in which crack
wave dynamics dominate. In two of the experiments described below, it 
was found that in the regime where $\zeta \sim 0.5$, 
the roughness amplitude is approximately 
independent of the {\it macroscopic average} crack velocity $V_m$ \cite{Depinning}. 
Only the crossover scale $\xi_c$, which is the upper limit of this regime, 
is found to depend on $V_m$. This suggests that the {\it local} crack velocity $V$ 
that enters Eq. (\ref{result-x}) may actually be roughly constant during 
localized depinning events and we will assume $V$ to be
a significant fraction of the Rayleigh velocity $c_R$.

Assuming that the viscoelastic broadening is the dominant effect (compared to
the scattering of the crack waves), one can write
$D \sim c^2 \tau _d$, such that Eq. (\ref{result-x})
finally reads (for $V \sim c$):
\be
\delta h \sim \xi_0  \sqrt{\frac{r_x}{c \tau_d}},\label{ofm}
\ee
assuming strongly disordered materials, for which $\sigma = O(1)$. 
It is reasonable to estimate the value of $\tau_d$ (which measures the 
viscoelastic lag between stresses and strains) as a typical vibrational
time $\tau_d \sim a/c$ where $a$ is an atomic distance. This leads to $\tau_d \sim
10^{-12}$ seconds.

\subsection{Stress corrosion of glass}

Four point bending experiments on soda-lime silica glass leading to stress
corrosion fracture were performed in a controled humidity environment. 
Typical values of the macroscopic average crack velocity $V_m$ are $V_m \sim
10^{-9} - 10^{-5}$ m/s. The regime where $\zeta = 0.5$ is found to extend between
$1$nm and tens of nanometers for the lowest velocities $V_m$. It is reasonable to
assume that in this material $\xi_0$ is of the order of the size of three to six
silica tetrahedra, i.e $\xi_0 \sim 1$nm, corresponding to the smallest 
scale of density fluctuations \cite{VanBruxel}. Taking $c \sim 2.\ 10^{3}$ m/s
and $\tau_d \sim 10^{-12}$ seconds, Eq. (\ref{result-x}) leads to 
$\delta h \sim 2$nm for $r_x = 10$nm, which corresponds well to observations 
in atomic force microscopy (AFM), (see \cite{Depinning} and Figure 1). 

The corresponding
value of the crossover length $\xi_L$ above which the effects of the
macroscopic geometry dominate, is larger than $10 \mu$m for a typical 
sample size of $L=1$ cm.Since this is much larger than the
crossover scale $\xi_c(V_m)$ 
separating the $\zeta=0.5$ from the $\zeta=0.8$ regime in this material, 
the finite sample size effects should not matter.

\subsection{Fatigue of a Ti$_3$Al-based alloy}

Fatigue experiments were carried out on compact tension specimens 
of a Ti$_3$Al-based alloy at a frequency of 30 Hz with a constant R-ratio of $0.1$.
Varying the maximum load allowed us to vary the average crack velocity 
again between $V_m = 10^{-9}$ and $10^{-5}$ m/s.   

This alloy contains faggots of needle shaped precipitates (of size 
$20/1 \mu$m) of the brittle 
$\alpha_2$ phase in the more ductile $\beta$ phase. The fracture mode was 
observed {\it in situ} using scanning electron microscopy (SEM). Cleavage cracks 
open in the $\alpha_2$ precipitates, blunt when extending into the $\beta$
matrix and finally coalesce together and with the main cracks.

The $\zeta=0.5$ regime is in this case 
observed at least down to $r \sim 10^{-2} \mu$m and up to $\xi_c = 10 \mu$m for the
lowest velocities $V_m$. It is reasonable to think that $\xi_0$ corresponds to the
size of heterogeneities contained within a needle, and hence significantly 
smaller than $1 \mu$m. Taking for $\xi_0$ the lower limit of the scaling region 
$10^{-2} \mu$m, one finds (with $c=5 \cdot 10^3$ m/s and $\tau_d = 10^{-12}$)
$\delta h \sim 0.5 \mu$m for $r_x \sim 10 \mu$m, which again concurs with experimental
findings \cite{Depinning} for which both AFM and SEM were used. Similar orders of magnitude
are found in the case of pure tension
fracture for which  $V_m$ is expected to be much larger \cite{PRE}.

The scale $\xi_L$ is again found to be tens of microns for $L=1$ cm, which is 
only of the same order of magnitude as 
the observed 
crossover scale $\xi_c(V_m)$ for the smallest $V_m$ studied in \cite{Depinning}.

\subsection{Fracture of ductile aluminium alloys}

Compact tension specimens of a ductile commercial aluminium alloy, 7010, were broken in fatigue
at a frequency of 10 Hz and a constant R ratio 0.1. Measured average crack velocities $V_m$
were ranging between $2 \cdot 10^{-9}$ m/s to $10^{-5}$ m/s. The largest value of the crossover length
$\xi_c(V_m)$ did not exceed $0.1 \mu$m in this case. Note that the values of $\xi_c(V_m)$ are
in this case always much smaller than the plastic zone size.

Taking again for $\xi_0$ the lower 
limit of the scaling region, i.e. $0.01 \mu$m and still $\tau_d \sim 10^{-12}$ s, 
one finds $\delta h \sim 0.02 \mu$m for
$r_x \sim 0.1 \mu$m whereas SEM observations \cite{Caramba} lead to $\delta h \sim 0.06 \mu$m for the same
$r_x$. The agreement is in this case more surprising since in this alloy the growth
of damage cavities should be dominated by plastic flow rather than crack front motion
and hence a longer
$\tau_d$ might be more reasonable. 
However, a recent direct study of these cavities using AFM \cite{Lil} has revealed a 
clearly anisotropic morphology,
where roughness amplitudes, in the direction where the exponent 0.5 is observed, are 
indeed similar to the ones measured at small length scales on fracture surfaces. 
Furthermore, the predicted orders of magnitude are also compatible with the results found 
on a rapidly quenched aluminium alloy of a different composition, in which the local porosity
resulting from the elaboration process might play an important role in the nucleation of 
damage cavities.

It would obviously be interesting to obtain more direct estimates of
$D$ (or equivalently $\tau_d$) and $\xi_0$ to check whether the above order of 
magnitudes in the
different materials are consistent. However, overall, the scenario in
which the small length scale exponent of $0.5$ is due to the existence of
diffusively broadened crack front waves in localized 
depinning events appears to be reasonable, 
at least in the more fragile samples. 

\section{Physical discussion} 

We have found that a model of diffusively damped crack front waves naturally leads
to a steady state roughening of fracture surfaces 
induced by the presence of random local
heterogeneities, with a roughness exponent $\zeta=0.5$, that could be the explanation for
part of the experimental data. However, there are many limitations to this result 
and complications that we expect on theoretical grounds. These we now discuss. 

\subsection{The low velocity limit} 

As is apparent from the linearized analysis discussed in this paper, the effects of the randomness 
become larger and larger at low velocities: see Eq. (\ref{result-x}). In addition, as shown by
Eq. (\ref{result-z}), lower velocities give rise to more and more anisotropy. 
Thus at some velocity, the linearized analysis will almost certainly breakdown. 
It is just such an apparent divergence in a linearized analysis that signals entry into the 
``critical'' regime in which the motion of the crack front changes qualitatively and becomes intermittent.
The non-linearities inherent in the dependence of the heterogeneities on the crack front position 
through the random function $\eta[x,Vt+f(x,t)]$ in Eq. (\ref{eqn-motion}) will then become important.   

It is instructive in this context to first consider the case that is best 
understood theoretically: the {\it in-plane} deformation of a crack front in the absence of 
elastic waves \cite{JRR1,RF-onset}.  In this case, a moving crack front has only logarithmic roughness
at long scales, 
as mentioned earlier. But at low crack velocities, this result obtains only for 
length scales longer than a correlation length $\xi(V)$ that diverges as a power of $V$. 
On smaller length scales, the physics is quite different, being dominated by the irregular 
start-stop motion characteristic of the non equilibrium dynamic 
critical-point at which the crack starts to advance.  
At the critical point, and for an advancing crack on length scales smaller than $\xi(V)$, 
the roughness of the crack front is determined by the avalanche-like processes by which the crack starts 
growing. The important non-linearities in this regime are those in the random dependence of 
the local fracture energy on the position of the crack front.  
These give rise to a critical crack-front roughness exponent predicted to be 
$\zeta_f \ge 1/3$, {\it larger} than that in the moving ``phase" ($\zeta_f=0$)
\cite{Krauth}.  
At this point, the effects of elastic waves on the onset of advance of planar cracks 
in randomly heterogeneous media are not understood, although some indications 
suggest that the critical behavior  may be similar to that in the absence of elastic waves 
\cite{schwarz-pulses}.\label{critical}

Similarly, in the case of primary interest to us, non planar crack front deformations, 
the whole concept of perturbing around a uniformly growing crack front is likely to 
lose its meaning at low velocities. This is because any given portion of the crack front is likely, 
as in the absence of crack waves, to spend most of its time essentially stationary, 
only occasionally advancing in a very jerky manner. 
This type of irregular local crack growth may well be incompatible with 
the propagation of corrugation waves. Nevertheless, another type of wave may well play an 
important role in the onset of crack growth and the roughness fracture surfaces at low \cite{RF-onset} 
velocities.  As discussed in \cite{RF-onset}, there are circumstances in which one might 
have shock waves of starting or stopping propagating along the crack front.  
If these involve a substantial non-planar component, then they would certainly affect the fracture surfaces.

A last problem is suggested by geometry: the fact that the angle of propagation of the 
corrugation waves --- if they do in fact still propagate at low crack velocities --- 
will be almost parallel to the crack front.  As the crack progresses, they and 
the sound and Rayleigh  waves associated with them will be reflected off 
the surfaces of the sample further complicating their effects. 

Since experiments report a $\zeta=0.5$ regime for rather small average crack velocities
(see section 4), 
with an amplitude that is, as mentioned above, found to be {\it independent} of the 
velocity \cite{Depinning}, one 
could 
argue that even though the macroscopic velocity $V_m$ is small, the instantaneous velocity 
during an `avalanche' in a strongly heterogeneous medium 
is a substantial fraction of the Rayleigh speed so that the present analysis 
might (at least
qualitatively) applicable.

\subsection{Non-linearities for rapidly advancing cracks}

As noted above, the corrugation wave induced roughness discussed here is similar to that 
for an over-damped elastic line driven through random impurities. For elastic lines, 
it is known that certain non-linearities can qualitatively change the behavior, 
including the roughness exponent \cite{PRL}. It would thus be
interesting to study in the context of fracture surfaces the effects of 
possible non-linearities. Preliminary indications are that for diffusively broadened 
crack corrugation waves traveling through a random medium, non-linearities that 
arise from deformations of the crack are {\it marginal} in the 
sense that one needs to go beyond a leading order perturbative analysis 
in the non-linearities to see whether they will alter the value of the roughness
exponent $\zeta$.\footnote{Because of the homogeneity of the elastodynamics equations,
the non linearities associated with deformations of the crack front will be of order $h^3/\lambda^3$ 
for wavelengths of order of $\lambda$. Non linearities of the 
Kardar-Parisi-Zhang type $(\partial_x h)^2$ or $(\partial_z h)^3$ are
known to become marginal in
$d=2$ dimensions \cite{BS}, whereas the front is a $d=1$ dimensional object. However,
in the presence of diffusively damped linear propagating waves, one can readily show that the `lower critical
dimension' where the non linearity is marginal is shifted from $d=2$ to $d=1$.}     
It is thus plausible that these non-linearities could give rise 
to the apparently universal exponent for fracture surface roughness.  
But whether or not this is the case,  the marginality suggests that whatever 
the correct asymptotic behavior, one expects on general grounds that one could observe the linear 
roughness exponent $\zeta=1/2$ over a substantial range of length scale before possibly crossing 
over to a different value, perhaps $\zeta=0.8$ on longer scales. 
[The crossover length is expected to be a
function of the amplitude of the non-linearity (a priori of order unity but there could be
small factors such as $\epsilon(V)$), the strength of the disorder $\sigma$, and 
the front velocity $V$.] It would be interesting (although a real technical challenge) to work out 
the theory in detail
and to decide whether or not the seductive scenario, where the exponent $\zeta=0.8$ is produced by 
the non-linear interaction of corrugation waves, is plausible. As argued below, however,
a perhaps more physically likely scenario for this crossover involves damage cavities and their coalescence.

Recent experiments on dynamic fracture of glass by Sharon and Fineberg
\cite{fineberg-waves} have observed crack front deformations caused by surface
imperfections that propagate for long distances.  These involve both small
amplitude corrugations -- with a shape qualitatively similar to figure 2
--- and much larger modulations of the crack velocity and the concomitant
in-plane deformation of the crack front.  Surprisingly, these pulses seem
to propagate without appreciable attenuation and have a shape of the
corrugations that is scale independent over more than an order of
magnitude in length scale.  The authors indicate that these and other
features of their experiments suggest that the pulses have a soliton-like
character indicative of the importance of non-linearities in spite of the
small amplitude of the corrugations.  Non-linear effects will in general
involve both in-plane and out-of-plane deformations. Via the existence of
small dimensionless numbers such as $\epsilon(V)$ and $(c_f-c_h)/c_R$,
where $c_f$ is the speed analogous to $c_R$ for in-plane waves
\cite{RF-planar,Rice3},
these could perhaps give rise to appreciable non-linear effects even with
small amplitude corrugations.  In any case, the observations provide clear
evidence for the existence of crack front waves and suggest that
non-linear interactions between them may be important at least in some
regimes.

\subsection{Long-range correlated heterogeneities}

Another effect that could change the roughness exponent of fracture surfaces 
is long-range correlations in the randomness that we have not 
considered so far.  
As discussed in reference \cite{REF}, correlations in 
the residual stresses in a material 
can affect the roughness of fracture surfaces by inducing random Mode II 
loading on the crack front as the crack 
grows and relieves the residual stresses.  If these frozen-in stresses have 
correlations that decay as a sufficiently 
small power law of distance, they will cause the surface to be rougher 
than it would otherwise have been.  
Sufficiently long range correlations will cause a positive $\zeta$ in the 
quasi-static case and a $\zeta>1/2$ in the presence of elastic waves.  
Whether long-range correlated residual stresses could by themselves cause the 
observed $\zeta \simeq 0.8$ is not clear, but, if this were 
indeed the correct cause, 
one would be left with the problem of understanding why similar power-law stress 
correlations are so ubiquitous. 

\subsection{In-plane crack front roughness}

It is tempting to attribute the observed crack front 
in-plane roughness exponent $\zeta_f$ of about 
$0.5$ to planar crack-front waves interacting with random impurities in a manner analogous 
to that analyzed in this paper for the corrugation waves. 
Evidence for propagation by localized
depinning has been obtained for fracture of a weakened interface between
plexiglass plates \cite{SchM}, although the average speed of events
between the recordings of crack front position, at 0.2 s intervals, is
less than about 50 mm/s, and we cannot be sure that the events are
actually dynamic in the sense of being inertially controlled.  The
problem with the dynamic depinning interpretation for in-plane
roughening is the effect of velocity dependent fracture energy on the crack front waves. Since the 
in-plane crack front waves change the local velocity of the front (at variance with the 
out-of-plane waves), any velocity dependence of the fracture energy will feed back into
the dynamics of these waves.  
As shown in \cite{RF-planar}, velocity strengthening damps the planar front-waves --- 
essentially by making their velocity complex --- while velocity weakening drives the 
crack front unstable at finite wavelengths.  Thus unless the velocity
strenghening is, fortuitously, extremely small (as is probably the case for glass
\cite{fineberg-waves}), or the crack somehow adjusts its velocity to a point of marginal stability, 
in a real material planar crack-front waves are unlikely to exist over a wide enough range of 
length scales to appreciably increase the roughness from the logarithmic behavior predicted 
for a moving crack with quasi-static dynamics. 

Since experiments that have measured crack front roughness have either been on stopped cracks or 
on --- at least apparently --- very slowly advancing cracks \cite{EBPD,Sch1,Sch2}, one would guess 
that the {\it critical} 
crack front roughness discussed in section (\ref{critical}) would be what is observed.  
It is somewhat puzzling, therefore, that the experiments have consistently observed front 
roughness exponents of order $0.5 - 0.65$ rather than $1/3$ (or even  
$\zeta_f \approx 0.39$, as suggested by a recent numerical simulation
\cite{Krauth}.) 
Note that a similar value has been 
found for the roughness of a slowly advancing contact line \cite{Rolley}, 
a problem that is expected to be 
in the same universality class as that of crack fronts since the quasi-static, 
linearized version of the two problems are the same.  A possibility, discussed in 
\cite{Schmitt-Vilotte}, is that the correlation length of the heterogeneities is substantial.
In the slowly moving regime, this could lead to a short length scale 
exponent of $\zeta_f=1/2$ that crosses-over
to $\zeta_f=1/3$ for large distances (see also \cite{Anusha}).

\section{From corrugation waves to coalescence of damage cavities ?}

As discussed above, the analysis we have done only makes sense when the moving crack front can support 
corrugation waves.  We have already raised some concerns about whether this will be the case 
at low velocities. 
But even in high velocity regimes in which these considerations do not play a role, one must certainly require that
the very concept of {\it a single} crack front makes sense. Various observations suggest that, in many materials, 
this may only be true at small enough length scales as the growth of cracks  
 in many complex materials (possibly including amorphous glassy 
materials) appears to occur by the
nucleation, growth and coalescence of damage cavities in the region surrounding 
the main crack front (see the discussion in \cite{Ravi-Chandar}). In such materials, 
the notion of a well defined
moving crack front only makes sense {\it locally}. One thus might expect that inside 
a growing cavity one would observe a
roughness exponent close to $0.5$ \cite{Lil}. This $0.5$ exponent could be due to the
mechanism discussed here which would be plausible if the local growth speed of the crack 
within a cavity were always fast -- or of a different origin, as discussed further 
in the conclusion. But in any case, on length scales larger than the typical
size of the cavities when they coalesce, one should observe a
crossover
to a new regime, dominated by inter-cavity correlations \cite{Pinault}. A natural supposition is that it is the 
physics of the formation and coalescence of the cavities which is responsible for the observed roughness 
exponent of $\zeta \simeq 0.8$ for which there is no theory at present. This scenario was proposed in \cite{Lil}, 
based on the observation of the roughness of growing cavities before coalescence in an aluminum alloy. 
Qualitatively similar ideas can also be found in \cite{Ravi-Chandar}. 

If one assumes that there is a nucleation rate $\gamma$
per unit time and unit length of new cavities ahead of the crack front, the
typical coalescence time of the cavities, $t_c$, is given by:
\be
\gamma \times  (V_c t_c)\times  t_c \sim 1,
\ee
where $V_c$ is the speed at which the cavities grow. The above equation means
that on the length scale $V_ct_c$ and time scale $t_c$ one cavity
will typically encounter another one; $Vt_c$ will be the size
of the coalescing cavities and thus the
crossover length $\xi_c$. Therefore:
\be
\xi_c \sim \sqrt{\frac{V_c}{\gamma}}.
\ee
Since $\gamma$ is expected to grow rapidly with the external stress, 
and therefore
with the macroscopic crack velocity $V_m$, this scenario could be
 compatible with the observed {\it decrease} of $\xi_c$ with 
increasing $V_m$.

In summary, we have argued that a crack-wave induced roughness
exponent $\zeta=0.5$ could hold over a range of length scales $r$, limited above by 
$r < \xi_c$, the scale of cavity coalescence. For very low macroscopic crack velocities $\xi_c$
may well exceed the maximum lengths observed, while for larger crack velocities, the crossover length $\xi_c$ would 
decrease into the observable range.

Needless to say, a statistical model based on the idea of cavity coalescence that would reproduce the correct
large scale value of $\zeta = 0.8$ is yet to be constructed. 

\section{Conclusion}

In this paper, we have reviewed some recent results concerning crack front waves. We have argued that out-of-plane
{\it corrugation} waves should strongly influence the roughness of fracture surfaces. The diffusive damping 
of these waves give rise to roughness exponents that are similar to those obtained in the Edwards-Wilkinson model,
although the underlying physics is very different. Our central prediction is that the roughness exponent is -- assuming
short range correlations in the disorder -- $\zeta=1/2$, except 
along two particular directions (those 
corresponding to the propagation of the crack front waves) where it is $1/4$. Furthermore, the roughness is predicted 
to be strongly anisotropic for cracks with instantaneous velocity much smaller than the Rayleigh speed. 
Although order of magnitudes on existing profiles are compatible with this
scenario, a direct experimental examination of these predictions would be very instructive.

We have discussed various limitations and complications that could obscure or 
even modify these results.  
We have speculated on the role of damage cavity coalescence, and the 
corresponding breakdown of the very concept of
a single crack {\it front}, to explain the still mysterious universal value 
$\zeta=0.8$ value of the roughness exponent at large length scales. 

Finally, we should mention that a value of $\zeta$ close to $0.5$ for fracture surfaces 
has also been found for minimal energy surfaces (see the discussion in \cite{Roux2,Batrouni}),
quasi-static scalar plasticity (discrete \cite{Batrouni} or 
continuous \cite{REF}) 
models and quasi-static vectorial discrete models \cite{Pietronero}, where the concept of crack
front waves is irrelevant. These models might be more relevant to explain the exponent 
$\zeta \sim 0.5$ found on fracture surfaces (or cavity surfaces) of
highly ductile, plastic materials where the instantaneous velocity is probably always small.
The fracture surfaces obtained in these models are expected to be isotropic, at variance 
with the above prediction based on corrugation waves. This feature should allow to distinguish 
the two mechanisms.

\vskip 1cm
Acknowledgments: J.P.B. wants to thank Harvard University for hospitality
during the period when this work was completed. Interesting discussions 
with J. Sethna are gratefully acknowledged. 
DSF thanks the National Science Foundation for support via DMR-9630064 and
DMR-9809334.
JRR thanks the Office of Naval Research for support via grant N00014-96-10777.


\begin{thebibliography}{99}


\bibitem{BS} for an introduction, see A. L. Barabasi, H. E. Stanley, `Fractal concepts in
Surface Growth', Cambridge University Press, 1995.

\bibitem{Batrouni} G. Batrouni, A. Hansen, {\it Fracture in three-dimensional fuse networks}, 
Phys. Rev. Lett. {\bf 80} (1998) 325 

\bibitem{EB1}  E. Bouchaud, G. Lapasset, J. Plan\`es, {\it Fractal dimension of
fracture surfaces: a universal value ?}, Europhys. Lett.,
{\bf 13}, (1990), 73.

\bibitem{PRL} J.-P. Bouchaud, E. Bouchaud, G.
Lapasset, J. Plan\`es, {\it Models of fractal cracks}, Phys. Rev. Lett. {\bf 71}, (1993) 2240.

\bibitem {Naveos} E. Bouchaud, S. Nav\'eos, {\it From quasi-static to rapid fracture}, 
J. Phys. I France {\bf 5}, (1995) 547; 

\bibitem{EB} For a review, see: E. Bouchaud, {\it Scaling properties of cracks},
J. Phys. Condensed Matter {\bf 9} (1997) 4319

\bibitem{Caramba} E. Bouchaud, M. Hinojosa, unpublished.

\bibitem{Claudin} see e.g.: P. Claudin, J.P. 
Bouchaud, M. E. Cates, J. Wittmer,  {\it Models of stress propagation in 
granular media}, Phys. Rev. E,
{\bf 57} (1998) 4441

\bibitem{Cot-Rice} B. Cottrell, J.R. Rice, {\it 
Slightly curved or kinked cracks}, Int. J. Frac. {\bf 16} (1980) 155 

\bibitem{EBPD} P. Daguier, E. Bouchaud, G. Lapasset, {\it Roughness of a crack front
pinned by microstructural obstacles}, Europhys. Lett. {\bf 31} (1995) 367.

\bibitem{PRE} P. Daguier, S. H\'enaux, E. Bouchaud, F. Creuzet, {\it Quantitative analysis of a fracture 
surface by Atomic Force Microscopy}, Phys. Rev. {\bf E53} (1996) 5637.

\bibitem{Depinning} P. Daguier, B. Nghiem, E. Bouchaud, F. Creuzet, {\it Pinning and depinning of crack
fronts in heterogeneous materials}, Phys. Rev. Lett., {\bf 78}, (1997) 1062.

\bibitem{Sch2} A. Delaplace, J. Schmittbuhl, K.J. M\aa l\o y, {\it 
High resolution description of a crack front in a heterogeneous Plexiglas 
block}, Phys. Rev. {\bf E60} (1999) 1337

\bibitem{DSF-unpub} D.S. Fisher, {\it On planar crack wave damping from viscoelasticity}, 
unpublished.

\bibitem{Pinault} A. Pinault, D. Francois, A. Zaoui, {\it Comportement 
m\'ecanique des mat\'eriaux}, Hermes, Paris (1995)

\bibitem{JRR2} H. Gao, J. R. Rice, {\it 
A first-order perturbation analysis of
crack trapping by arrays of obstacles}, J. Appl. Mech. {\bf 56} (1989) 828 

\bibitem{Rolley} C. Guthmann, R. Gombrowicz, V. Repain, E. Rolley, Phys. Rev. Lett, {\it Roughness of the Contact Line on a Disordered Substrate}
{\bf 80} (1998) 2865

\bibitem{Roux2} A. Hansen, E. L. Hinrichsen, S. Roux, Phys. Rev. Lett,
{\it Roughness of crack interfaces} {\bf 66} (1991) 2476 

\bibitem{Anusha} A. Hazareesing, M. M\'ezard, 
{\it Wandering of a contact line at thermal equilibrium}, Phys. Rev. {\bf E 60} (1999) 1269


\bibitem{Sethna} J. Hodgdon, J.P. Sethna, {\it 
Derivation of a general three-dimensional crack-propagation law: A 
generalization of the principle of local symmetry}, Phys. Rev. {\bf B}
{\bf 47} (1993) 4831.

\bibitem{Hull} D. Hull, P. Beardmore, {\it Velocity of propagation of
cleavage cracks in tungsten}, Int. J. Frac. {\bf 2} (1966) 468

\bibitem{Krauth} A. Rosso, W. Krauth, {\it Roughness at
the depinning threshold for a long-range elastic string}, e-print cond-mat/0107527

\bibitem{log-expts} H. Larralde, R. Ball, {\it The shape of slowly growing cracks},
 Europhys. Lett. {\bf 30} (1985) 287

\bibitem{Roux} K.J. M\aa l\o y, A.
Hansen, E. L. Hinrichsen, S. Roux, {\it Experimental measurements of the roughness
of brittle cracks}, Phys. Rev. Lett., {\bf 68}, 213,
1992.

\bibitem{SchM} K. M\aa l\o y and J. Schmittbuhl, {\it Dynamical
events during slow crack propagation}, Phys. Rev. Lett., in press, 2001


\bibitem{Mandelbrot} B.B. Mandelbrot, D.E. Passoja, A.J. Paullay, {\it Fractal character of 
fracture surfaces of metals}, 
Nature (London), {\bf 308}, 721, 1984

\bibitem{Rice2} J. W. Morrissey, J. R. Rice, {\it Crack front waves}, J. Mech. Phys. Solids {\bf 46} (1998) 467

\bibitem{Rice3} J. W. Morrissey, J. R. Rice, {\it 
Perturbative simulations of
crack front waves}, J. Mech. Phys. Solids {\bf 48} (2000) 122 

\bibitem{Pietronero} A. Parisi, G. Caldarelli, L. Pietronero, {\it
Roughness of fracture surfaces}, e-print cond-mat/0004374

\bibitem{Lil} F. Paun, E. Bouchaud, {\it Morphology of damage cavities}, in preparation.

\bibitem{Rice4} G. Perrin, J. R. Rice, {\it 
Disordering of a dynamic planar crack
front in a model elastic medium of randomly variable toughness}, J. Mech. Phys. Solids {\bf 42}
(1994) 1047


\bibitem{RF-planar} S. Ramanathan, D.S. Fisher, {\it 
Dynamics and instabilities of
planar tensile cracks in  heterogeneous media}, Phys. Rev. Lett. {\bf 79} (1997) 877

\bibitem{REF} S. Ramanathan, D. Erta\c s, and D. S. Fisher, {\it Quasi-static crack
propagation in heterogeneous media}, Phys. Rev. Lett. {\bf 79} (1997) 873

\bibitem{RF-onset} 
S. Ramanathan and D. S. Fisher, {\it 
Onset of Propagation of Planar Cracks in Heterogeneous Media}, 
Phys. Rev {\bf B 58}, 6026 (1998)

\bibitem{RF-NP} S. Ramanathan and D.S. Fisher, {\it Corrugation waves in dynamic fracture}, 
in preparation

\bibitem{Ravi-Chandar} K. Ravi-Chandar and B. Yang, {\it 
On the role of microcracks in the
dynamic fracture of brittle materials}, Journal of Mechanics and Physics of Solids, 
{\bf 45}, (1997), 535-563.

\bibitem{JRR1} J. R. Rice, {\it First order variation in elastic fields due to
variation in location of a planar crack front}, J. Appl. Mech. {\bf 52} (1985) 571 

\bibitem{SRVM} J. Schmittbuhl, S. Roux, J.P. Vilotte, K. J. M\aa l\o y, {\it Interfacial crack
pinning: effect of non local interactions},
Phys. Rev. Lett. {\bf 74} (1995) 1787.

\bibitem{Sch1} J. Schmittbuhl, K.J. M\aa l\o y, {\it Direct observation of a self-affine
crack propagation}, Phys. Rev. Lett. {\bf 78} (1997) 3888

\bibitem{Schmitt-Vilotte} J. Schmittbuhl, J.P. Vilotte, {\it 
Interfacial crack front
wandering: influence of quenched noise correlations}, Physica A{\bf 270} (1999) 42

\bibitem{schwarz-pulses} J. Schwarz and D.S. Fisher {\it Effects of stress pulses on depinning transitions in elastic media}, in preparation

\bibitem{fineberg-waves} E. Sharon, G. Cohen, J. Fineberg, 
{\it Propagating solitary waves
along a rapidly moving crack front}, Nature, {\bf 410} (2001) 68

\bibitem{VanBruxel} L. Van Brutzel, {\it Contribution \'a l'\'etude des m\'ecanismes 
de rupture dans les amorphes}, Ph. D. Dissertation, Paris VI, 1999, unpublished.


\bibitem{Wallner-expts} H. Wallner, {\it Linienstrukturen an bruchfl\"achen}, 
Z. Physik, {\bf 114} (1939) 368

\bibitem{Mov-Will1} J. R. Willis, A. B. Movchan, {\it Dynamic weight functions for a
moving crack. I. Mode I loading}, J. Mech. Phys. Solids, {\bf 43} (1995) 319

\bibitem{Mov-Will2}  J. R. Willis, A. B. Movchan, {\it 
Three-dimensional dynamic
perturbation of a propagating crack}, J. Mech. Phys. Solids, {\bf 45} (1997) 591



\end{thebibliography}
\end{document}